 \documentclass[mlabstract]{jmlr}

\usepackage{longtable}

\usepackage{booktabs}
\usepackage[load-configurations=version-1]{siunitx} 

\theorembodyfont{\upshape}
\theoremheaderfont{\scshape}
\theorempostheader{:}
\theoremsep{\newline}

\usepackage{mathtools}
\usepackage{amsmath,amsfonts,amssymb,algorithm,algorithmicx,algpseudocode,graphicx}
\usepackage{xcolor}
\definecolor{cadmiumorange}{rgb}{0.93, 0.53, 0.18}
\definecolor{darkorange}{rgb}{1.0, 0.55, 0.0}
\definecolor{mangotango}{rgb}{1.0, 0.51, 0.26}
\definecolor{darkpurple}{rgb}{0.5,0,0.5}
\definecolor{mydarkgrey}{rgb}{0.27, 0.27, 0.27}
\definecolor{antiquefuchsia}{rgb}{0.57, 0.36, 0.51}
\definecolor{cadetblue}{rgb}{0., 0.62, 0.63}
\definecolor{brightcerulean}{rgb}{0.11, 0.67, 0.84}
\definecolor{brightmaroon}{rgb}{0.76, 0.13, 0.28}
\definecolor{mymaroon}{rgb}{0.62, 0.05, 0.19}
\definecolor{mymaroon2}{cmyk}{0, 0.739, 0.614, 0.368}
\definecolor{darkpurple}{rgb}{0.5,0,0.5}
\definecolor{mydarkgrey}{rgb}{0.27, 0.27, 0.27}
\definecolor{mymagenta}{rgb}{0.8125, 0, 0.8125}
\definecolor{mygreen}{rgb}{0.37, 0.62, 0.63}
\definecolor{cambridgeblue}{rgb}{0.64, 0.76, 0.68}
\definecolor{carmine}{rgb}{0.59, 0.0, 0.09}
\usepackage{nicefrac}   
\usepackage{graphicx} 
\usepackage{bibunits}
\usepackage{indentfirst}
\usepackage{url}
\newcommand{\de}{\text{d}}
\hypersetup{colorlinks,urlcolor=cadetblue, citecolor=mygreen, linkcolor=carmine}
\usepackage{mathtools,amsmath}
\usepackage{url}
\usepackage{bibunits}

\jmlrvolume{}
\firstpageno{1}
\editors{List of editors' names}

\jmlryear{2025}
\jmlrworkshop{Symmetry and Geometry in Neural Representations}

\title[Meta-learning three-factor rules]{Meta-learning three-factor plasticity rules for\\ structured credit assignment with sparse feedback 
}

  \author{\Name{Dimitra Maoutsa} \Email{dimitra.maoutsa@gmail.com}\\
  }

\begin{document}

\maketitle

\begin{abstract}
Biological neural networks learn complex behaviors from sparse, delayed feedback using local synaptic plasticity, yet the mechanisms enabling structured credit assignment remain elusive. In contrast, artificial recurrent networks solving similar tasks typically rely on biologically implausible global learning rules or hand-crafted local updates. The space of local plasticity rules capable of supporting learning from delayed reinforcement remains largely unexplored. Here, we present a meta-learning framework that discovers local learning rules for structured credit assignment in recurrent networks trained with sparse feedback. Our approach interleaves local neo-Hebbian-like updates during task execution with an outer loop that optimizes plasticity parameters via \textbf{tangent-propagation through learning}. The resulting three-factor learning rules enable long-timescale credit assignment using only local information and delayed rewards, offering new insights into biologically grounded mechanisms for learning in recurrent circuits.
\end{abstract}

\section{Introduction}
\label{sec:intro}

Learning in biological organisms involves changes in synaptic connections (synaptic plasticity) between neurons~\citep{bailey1993structural, mayford2012synapses}. Synaptic changes are believed to underlie memory formation and are essential for adaptive behaviour~\citep{hopfield1982neural}. Experimental evidence suggests that synaptic changes depend on the co-activation of pre- and postsynaptic activity~\citep{bi1998synaptic, sjostrom2001rate}, and possibly other local variables available at the synaptic site~\citep{graupner2012calcium, pedrosa2020voltage}. These unsupervised synaptic modifications have explained activity-dependent circuit refinement during development such as the emergence of functional properties like receptive field formation based on naturalistic input statistics~\citep{martin2000synaptic,blais1997receptive,brito2016nonlinear,gutig2003learning,law1994formation}.

Yet, most organisms routinely solve complex tasks that require feedback through explicit supervisory or reinforcement signals. These signals are believed to gate or modulate plasticity, acting in the form of a third factor that scales and also possibly imposes the direction of the required synaptic modifications~\citep{kusmierz2017learning, sosis2024distinct} to facilitate long-lasting alignment of representations to behaviourally relevant dimensions~\citep{benezra2024learning}. How error- or reward-related information is propagated through the recurrent interactions is not yet clear.
  While prior work has largely focused on hand-crafted synaptic updates for unsupervised neural circuit self-organization, or biologically plausible approximations of backpropagation~\citep{miconi2018differentiable}, the space of plasticity rules capable of supporting structured credit assignment from delayed feedback remains vastly underexplored.

\begin{figure}[h!]
\includegraphics[width=\linewidth]{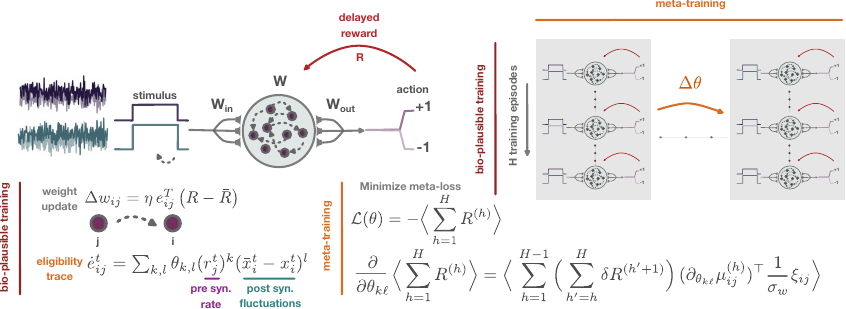}
\caption{\textbf{Outline of the proposed meta-learning framework.}} 
\label{fig:1}
\end{figure}

Backpropagation through time (BPTT), the standard approach for training recurrent neural networks (RNNs), is biologically implausible since it requires symmetric forward and backward connections and non-local information~\citep{lillicrap2016random, guerguiev2017towards}. Although recent work has reformulated BPTT into more biologically plausible variants using random feedback~\citep{lillicrap2016random, murray2019local}, truncated approximations, or by learning feedback pathways~\citep{lindsey2020learning,shervani2023meta}, these methods require continuous error signals to refine recurrent connections.

Here, we adopt a bottom-up approach: instead of imposing hand-designed synaptic rules, we discover biologically plausible plasticity rules that support learning through delayed reinforcement signals via meta-optimisation~\citep{schmidhuber1996simple}. Building on recent work~\citep{confavreux2023meta}, we parameterise plasticity rules as functions of local signals (presynaptic activity, postsynaptic activity, and synapse size) and meta-learn their parameters within a second reinforcement learning loop.
With that, our ongoing work tackles the following questions:
\begin{itemize}
	\item	\textbf{Which local learning rules can implement structured credit assignment under biological constraints?}
	\item	\textbf{
    Do different plasticity rules give rise to different representations and/or dynamics?}
	\end{itemize}

Here we demonstrate that different forms of plasticity naturally lead to qualitatively different learning trajectories and internal representations, akin to their gradient-based counterparts trained with different learning rules.

\section{Method}

\paragraph{Network dynamics.} We consider recurrent neural networks of firing rate neurons coupled through a synaptic matrix ${\mathbf{W} \in \mathcal{R}^{N\times N}}$~\citep{sompolinsky1988chaos, barak2017recurrent}, with additional input and output matrices ${\mathbf{W}_{\text{in}} \in \mathcal{R}^{N\times N_{\text{in}}}}$ and ${\mathbf{W}_{\text{out}} \in \mathcal{R}^{N_{\text{out}}\times N }}$ that route task-relevant input into the recurrent circuit and read out network activation to generate task-specific outputs (actions). The equations governing the network dynamics are
\begin{align}\label{main-1}
\frac{\de \mathbf{x}^t}{\de t} &= -\mathbf{x}^t + \mathbf{W} \boldsymbol{\phi}({\mathbf{x}}^t) + \mathbf{W}_{\text{in}} \mathbf{u}^t,\quad \quad
\mathbf{r}^t = \boldsymbol{\phi}({\mathbf{x}}^t) \dot{=} \tanh{({\mathbf{x}}^t)},
\end{align}
where $\mathbf{x}^t \in \mathcal{R}^N$ is the vector of pre-activations (or input currents) to each neuron in the network,  $\boldsymbol{\phi}(\cdot):\mathcal{R}^N \rightarrow \mathcal{R}^N $ denotes the single-neuron transfer functions, $\mathbf{r}^t \in \mathcal{R}^N$ is the vector of instantaneous firing rates, $\mathbf{u}^t$ stands for the activity of the $N_{\text{in}}$ input neurons. In the terms above, the $\cdot^t$ superscript indicates time dependence.
Network outputs $\mathbf{z}^t$ are obtained from linear read-out neurons as
  $  \mathbf{z}^t = \mathbf{W}_{\text{out}} \mathbf{r}^t.$

\paragraph{Sparse feedback and parametrized learning rules.} We consider networks that learn cognitive tasks using biologically plausible local learning rules, guided by sparse reinforcement signals $R$ provided only at the end of each training episode.
Each synapse between a pre-synaptic unit $j$ and a post-synaptic unit $i$ maintains an eligibility trace $e_{ij}$~\citep{izhikevich2007solving}, which integrates the history of (co-)activation during the episode. We define the evolution of eligibility traces with differential equations of the form
\begin{equation}\label{eq:eligi}
\frac{\de e^t_{ij}}{\de t} = \mathcal{H}_\theta( r^t_j, x^t_i) - \frac{e^t_{ij}}{\tau_e}= \sum_{0\leq k;\ell\leq d;}{ \theta_{k,\ell} \; \left(r^t_j\right)^k  \left(\bar{x}_i - x^t_i\right)^\ell  } - \frac{e^t_{ij}}{\tau_e}, 
\end{equation}
where $\tau_e$ is a decay time-scale, $\bar{x}_i$ is a running average of the pre-activation of neuron $i$, and $\theta_{k,\ell} \in \mathcal{R}$ are learnable coefficients.
In contrast to eligibility traces based solely on first-order correlations~\citep{gerstner2018eligibility}, we use here a polynomial expression that captures richer interactions between pre- and post-synaptic activity.
Each coefficient $\theta_{k,\ell}$ can be construed as a term-specific learning rate, which may be positive (Hebbian), negative (anti-Hebbian), or zero. 
In our experiments, we set $d = 5$.

The recurrent weight matrix $\mathbf{W}$ gets updated at the end of each training episode according to a reward-modulated learning rule
\begin{equation} \label{eq:update_rule}
\pi_{\Theta}\left(  \mathbf{\Delta W}^{(h)} \mid \boldsymbol{\Theta} \right) = \mathcal{MN}\left( \boldsymbol{\mu_{\Theta}}^{(h)},\, \sigma^2_w \, \mathbf{I}_N, \mathbf{I}_N \right) \quad \text{with}\quad  [\boldsymbol{\mu_{\Theta}}^{(h)}]_{ij} = \eta \, e^{T_h}_{ij}\, \left( {R}^{(h)} - \bar{R}^{(h)}\right) ,
\end{equation}
where with $\mathcal{MN}\left(\boldsymbol{\mu}, \boldsymbol{\Sigma}, \mathbf{V}\right)$ we denote the matrix normal distribution with mean $\boldsymbol{\mu} \in \mathcal{R}^{N\times N}$, and $\boldsymbol{\Sigma}$ and $\mathbf{V}$ the positive semi-definite matrices are the row- and column-variance, while superscript $h$ indicates episode index,
$\eta$ denotes the learning rate,
$e^{T_s}_{ij}$ stands for the eligibility trace accumulated till the end of the $h$-th episode $T_h$, while $R$, $\bar{R}$ stand for the obtained and the expected reward. Here, we model reward expectations for each type of trial independently as a running average of past rewards for this trial type~\citep{miconi2017biologically}. This update rule enables credit assignment through the interaction between synaptic eligibility and trial-specific reward prediction error, consistent with neo-Hebbian three-factor learning rules hypothesized to operate in biological circuits~\citep{gerstner2018eligibility}. 

\paragraph{Meta-learning plasticity rules.} While previous work has relied on hand-crafted eligibility trace dynamics and synaptic update rules to train recurrent neural networks with sparse feedback~\citep{miconi2017biologically}, we instead adopt a meta-learning approach to learn the parameters of the plasticity rules.
Our framework consists of two nested training loops:
\textcolor{mymaroon}{\textbf{(i)}} an inner loop in which the recurrent network is trained over several episodes using local learning rules and sparse reinforcement signals provided at the end of each episode \textcolor{carmine}{\textbf{(bio-plausible training)}}, as described above; and
\textcolor{cadmiumorange}{\textbf{(ii)}} an outer loop that optimizes the plasticity meta-parameters $\boldsymbol{\Theta} = \{ \{\theta_{k,\ell}\}^5_{k,\ell=0}\, \}$ via gradient descent using \textbf{tangent-propagation through learning} (forward-mode differentiation through learning) on a meta-loss computed over $H$ training episodes (trials) \textcolor{cadmiumorange}{\textbf{(meta-training)}}.
This approach allows the learning rules themselves to be adapted to the task, rather than be fixed a priori.

\paragraph{Tangent-propagation through learning.} 
Our goal is to optimise the learning rule parameters $\boldsymbol{\Theta}$ to maximise task performance, quantified as the expected cumulative reward $\langle \sum_h R \rangle$ obtained after a fixed number of learning episodes.
However, the reward $R$ depends on the network’s output, which is determined by synaptic weights $\mathcal{W} = \{\mathbf{W}_{in}, \mathbf{W}, \mathbf{W}_{out}\}$, with $\mathbf{W}$ evolving under the update rule (Eq.\ref{eq:update_rule}). 
Since weights depend on eligibility traces $e_{ij}$, themselves parameterised by $\boldsymbol{\Theta}$, the reward depends on the plasticity parameters through $\mathcal{W}$ and $\boldsymbol{\Theta}$. 
Directly computing $\nabla_{\boldsymbol{\Theta}} \langle \sum_h R \rangle$ by backpropagating through the learning dynamics is computationally prohibitive since learning requires several hundreds of trials~\citep{lindsey2020learning}.
We therefore, here, adopt a REINFORCE-inspired estimator~\citep{williams1992simple}, which involves computing the gradient of an expected value by observing outcomes (\emph{the rewards}) and scaling a measure of what elicited that outcome (\emph{the weight updates}) with the associated reward. 
Thus, we approximate the gradient of the expected reward by
\begin{equation}\label{appeq:reinforce-gradient}
\nabla_\Theta \langle \sum_h  R^{(h)} \rangle \approx \langle \sum_h  \sum^H_{h^\prime=h+1} R^{(h^\prime)}  \nabla_\Theta \log \pi(\boldsymbol{\Delta}\mathbf{W}^{(h)} \mid \boldsymbol{\Theta}) \rangle \approx  \langle \sum_{h}  \sum^H_{h^\prime=h+1}  (R^{(h^\prime)} - \bar{R}^{(h^\prime)}) \;\,\nabla_\Theta \log \pi(\boldsymbol{\Delta}\mathbf{W}^{(h)}  \mid \boldsymbol{\Theta}) \rangle,
\end{equation}
where $\bar R$ stands for the baseline reward, and thus $(R - \bar{R})$ denotes the reward prediction error.
Introducing the expression of the plasticity rule, we have in a component-wise formulation for each dimensional component of the plasticity parameters $\boldsymbol{\Theta}$
\begin{align}\label{appeq:dimensional-gradient}
\frac{\partial}{\partial \theta_{k,\ell}} \Big\langle \sum_{h=1}^{H-1} R^{(h)}\Big \rangle
&= \Bigg\langle \mathlarger\sum_{h=1}^{H-1} \Big(\sum_{h^\prime=h}^{H-1}\delta R^{(h^\prime+1)} \Big)\, \frac{1}{\sigma_w^{2}}\sum_{i=1}^{N}\sum_{j=1}^{N}
   \Big(\Delta w^{\,(h)}_{ij}-\mu^{\,(h)}_{ij}\Big)\;
   \frac{\partial \mu^{\,(h)}_{ij}}{\partial \theta_{k, \ell}} \Bigg\rangle_S,
  \end{align}
where the expectation $\langle \cdot \rangle_S$ is considered over independent sessions $S$. This requires the computation of the sensitivity of the mean weight update wrt to the plasticity parameters $ \frac{\partial \mu^{\,(h)}_{ij}}{\partial \theta_{kl}}$ over training. To that end, we propagate the gradients of the within-trial pre-activations $\mathbf{x}^t$, $\boldsymbol{\chi}_{k,\ell}^t \in \mathcal{R}^N$ (\textbf{state tangent}), the pre-activation trace $\bar{\mathbf{x}}^t$, $\boldsymbol{\psi}_{k,\ell}^t \in \mathcal{R}^N$ (\textbf{trace tangent}), and of the eligibility traces of each synaptic pair $ij$, $e^t_{ij}$, ${[\mathbf{z}^t_{k,\ell}]}_{ij}$ (\textbf{eligibility tangent}), as well as inter-trial sensitivities of weight matrices (\textbf{weight matrix tangent}), $\mathbf{U}_{k,\ell}^{(h)}$ (c.f. Appendix Sec.~\ref{apd:tangents}).

\newpage





\bibliography{pmlr-sample}

@article{sosis2024distinct,
  title={Distinct dopaminergic spike-timing-dependent plasticity rules are suited to different functional roles},
  author={Sosis, Baram and Rubin, Jonathan E},
  journal={{bioRxiv}},
  year={2024}
}

@article{murphy2009balanced,
  title={Balanced amplification: a new mechanism of selective amplification of neural activity patterns},
  author={Murphy, Brendan K and Miller, Kenneth D},
  journal={{N}euron},
  volume={61},
  number={4},
  pages={635--648},
  year={2009},
  publisher={Elsevier}
}

@article{christodoulou2022regimes,
  title={Regimes and mechanisms of transient amplification in abstract and biological neural networks},
  author={Christodoulou, Georgia and Vogels, Tim P and Agnes, Everton J},
  journal={{PLoS Computational Biology}},
  volume={18},
  number={8},
  pages={e1010365},
  year={2022},
  publisher={{Public Library of Science San Francisco, CA USA}}
}

@article{barak2017recurrent,
  title={Recurrent neural networks as versatile tools of neuroscience research},
  author={Barak, Omri},
  journal={{Current {O}pinion in {N}eurobiology}},
  volume={46},
  pages={1--6},
  year={2017},
  publisher={{Elsevier}}
}

@article{lindsey2020learning,
  title={Learning to learn with feedback and local plasticity},
  author={Lindsey, Jack and Litwin-Kumar, Ashok},
  journal={{Advances in Neural Information Processing Systems}},
  volume={33},
  pages={21213--21223},
  year={2020}
}

@article{williams1992simple,
  title = {Simple statistical gradient-following algorithms for connectionist reinforcement learning},
  author = {Williams, Ronald J},
  journal = {{Machine Learning}},
  volume = {8},
  number = {3-4},
  pages = {229--256},
  year = {1992},
  publisher = {{Springer}}
}

@inproceedings{miconi2018differentiable,
  title={Differentiable plasticity: training plastic neural networks with backpropagation},
  author={Miconi, Thomas and Stanley, Kenneth and Clune, Jeff},
  booktitle={{International Conference on Machine Learning}},
  pages={3559--3568},
  year={2018},
  organization={{PMLR}}
}

@article{gerstner2018eligibility,
  title={Eligibility traces and plasticity on behavioral time scales: experimental support of neohebbian three-factor learning rules},
  author={Gerstner, Wulfram and Lehmann, Marco and Liakoni, Vasiliki and Corneil, Dane and Brea, Johanni},
  journal={{Frontiers in Neural Circuits}},
  volume={12},
  pages={53},
  year={2018},
  publisher={{Frontiers Media SA}}
}

@article{sompolinsky1988chaos,
  title={Chaos in random neural networks},
  author={Sompolinsky, Haim and Crisanti, Andrea and Sommers, Hans-Jurgen},
  journal={{Physical Review Letters}},
  volume={61},
  number={3},
  pages={259},
  year={1988},
  publisher={{APS}}
}

@article{izhikevich2007solving,
  title={Solving the distal reward problem through linkage of STDP and dopamine signaling},
  author={Izhikevich, Eugene M},
  journal={{Cerebral Cortex}},
  volume={17},
  number={10},
  pages={2443--2452},
  year={2007},
  publisher={{Oxford University Press}}
}

@article{miconi2017biologically,
  title={Biologically plausible learning in recurrent neural networks reproduces neural dynamics observed during cognitive tasks},
  author={Miconi, Thomas},
  journal={{Elife}},
  volume={6},
  pages={e20899},
  year={2017},
  publisher={{eLife Sciences Publications, Ltd}}
}

@article{lillicrap2016random,
  title={Random synaptic feedback weights support error backpropagation for deep learning},
  author={Lillicrap, Timothy P and Cownden, Daniel and Tweed, Douglas B and Akerman, Colin J},
  journal={{Nature Communications}},
  volume={7},
  number={1},
  pages={13276},
  year={2016},
  publisher={{Nature Publishing Group UK London}}
}

@article{guerguiev2017towards,
  title={Towards deep learning with segregated dendrites},
  author={Guerguiev, Jordan and Lillicrap, Timothy P and Richards, Blake A},
  journal={{Elife}},
  volume={6},
  pages={e22901},
  year={2017},
  publisher={{eLife Sciences Publications, Ltd}}
}

@article{mayford2012synapses,
  title={Synapses and memory storage},
  author={Mayford, Mark and Siegelbaum, Steven A and Kandel, Eric R},
  journal={{Cold Spring Harbor perspectives in biology}},
  volume={4},
  number={6},
  pages={a005751},
  year={2012},
  publisher={{Cold Spring Harbor Lab}}
}

@article{bailey1993structural,
  title={Structural changes accompanying memory storage.},
  author={Bailey, Craig H and Kandel, Eric R},
  journal={{Annual Review of Physiology}},
  year={1993},
  publisher={{Annual Reviews}}
}

@article{schmidhuber1996simple,
  title={Simple principles of metalearning},
  author={Schmidhuber, Juergen and Zhao, Jieyu and Wiering, MA},
  journal={{Technical report IDSIA}},
  volume={69},
  pages={1--23},
  year={1996},
  publisher={{IDSIA}}
}

@article{shervani2023meta,
  title={Meta-learning biologically plausible plasticity rules with random feedback pathways},
  author={Shervani-Tabar, Navid and Rosenbaum, Robert},
  journal={Nature {C}ommunications},
  volume={14},
  number={1},
  pages={1805},
  year={2023},
  publisher={{Nature Publishing Group UK London}}
}

@article{confavreux2023meta,
  title={Meta-learning families of plasticity rules in recurrent spiking networks using simulation-based inference},
  author={Confavreux, Basile and Ramesh, Poornima and Goncalves, Pedro J and Macke, Jakob H and Vogels, Tim},
  journal={{Advances in Neural Information Processing Systems}},
  volume={36},
  pages={13545--13558},
  year={2023}
}

@article{hopfield1982neural,
  title={Neural networks and physical systems with emergent collective computational abilities.},
  author={Hopfield, John J},
  journal={{Proceedings of the National Academy of Sciences}},
  volume={79},
  number={8},
  pages={2554--2558},
  year={1982}
}

@article{graupner2012calcium,
  title={Calcium-based plasticity model explains sensitivity of synaptic changes to spike pattern, rate, and dendritic location},
  author={Graupner, Michael and Brunel, Nicolas},
  journal={{Proceedings of the National Academy of Sciences}},
  volume={109},
  number={10},
  pages={3991--3996},
  year={2012},
  publisher={National Academy of Sciences}
}

@article{pedrosa2020voltage,
  title={Voltage-based inhibitory synaptic plasticity: network regulation, diversity, and flexibility},
  author={Pedrosa, Victor and Clopath, Claudia},
  journal={bio{R}xiv},
  pages={2020--12},
  year={2020}
}

@article{bi1998synaptic,
  title={Synaptic modifications in cultured hippocampal neurons: dependence on spike timing, synaptic strength, and postsynaptic cell type},
  author={Bi, Guo-qiang and Poo, Mu-ming},
  journal={{Journal of {N}euroscience}},
  volume={18},
  number={24},
  pages={10464--10472},
  year={1998},
  publisher={{Society for Neuroscience}}
}

@article{sjostrom2001rate,
  title={Rate, timing, and cooperativity jointly determine cortical synaptic plasticity},
  author={Sj{\"o}str{\"o}m, Per Jesper and Turrigiano, Gina G and Nelson, Sacha B},
  journal={Neuron},
  volume={32},
  number={6},
  pages={1149--1164},
  year={2001},
  publisher={Elsevier}
}

@article{lemieux2014control,
  title={Control variates},
  author={Lemieux, Christiane},
  journal={{Wiley StatsRef: Statistics Reference Online}},
  pages={1--8},
  year={2014},
  publisher={{Wiley Online Library}}
}

@article{kusmierz2017learning,
  title={Learning with three factors: modulating Hebbian plasticity with errors},
  author={Ku{\'s}mierz, {\L}ukasz and Isomura, Takuya and Toyoizumi, Taro},
  journal={{Current Opinion in Neurobiology}},
  volume={46},
  pages={170--177},
  year={2017},
  publisher={Elsevier},
}

@article{martin2000synaptic,
  title={Synaptic plasticity and memory: an evaluation of the hypothesis},
  author={Martin, Stephen J and Grimwood, Paul D and Morris, Richard GM},
  journal={{Annual Review of Neuroscience}},
  volume={23},
  number={1},
  pages={649--711},
  year={2000},
  publisher={{Annual Reviews}}
}

@article{murray2019local,
  title={Local online learning in recurrent networks with random feedback},
  author={Murray, James M},
  journal={Elife},
  volume={8},
  pages={e43299},
  year={2019},
  publisher={{eLife Sciences Publications, Ltd}}
}

@article{brito2016nonlinear,
  title={Nonlinear Hebbian learning as a unifying principle in receptive field formation},
  author={Brito, Carlos SN and Gerstner, Wulfram},
  journal={{PLoS computational biology}},
  volume={12},
  number={9},
  pages={e1005070},
  year={2016},
  publisher={{Public Library of Science}}
}

@article{law1994formation,
  title={Formation of receptive fields in realistic visual environments according to the Bienenstock, Cooper, and Munro (BCM) theory.},
  author={Law, C Charles and Cooper, Leon N},
  journal={{Proceedings of the National Academy of Sciences}},
  volume={91},
  number={16},
  pages={7797--7801},
  year={1994}
}

@article{asllani2018topological,
  title={Topological resilience in non-normal networked systems},
  author={Asllani, Malbor and Carletti, Timoteo},
  journal={{Physical Review E}},
  volume={97},
  number={4},
  pages={042302},
  year={2018},
  publisher={{APS}}
}

@article{asllani2018structure,
  title={Structure and dynamical behavior of non-normal networks},
  author={Asllani, Malbor and Lambiotte, Renaud and Carletti, Timoteo},
  journal={{Science {A}dvances}},
  volume={4},
  number={12},
  pages={eaau9403},
  year={2018},
  publisher={{American Association for the Advancement of Science}}
}

@article{benezra2024learning,
  title={Learning enhances behaviorally relevant representations in apical dendrites},
  author={Benezra, Sam E and Patel, Kripa B and Campos, Citlali P{\'e}rez and Hillman, Elizabeth MC and Bruno, Randy M},
  journal={Elife},
  volume={13},
  pages={RP98349},
  year={2024},
  publisher={{eLife Sciences Publications Limited}}
}

@article{gutig2003learning,
  title={Learning input correlations through nonlinear temporally asymmetric Hebbian plasticity},
  author={G{\"u}tig, Robert and Aharonov, Ranit and Rotter, Stefan and Sompolinsky, Haim},
  journal={{Journal of Neuroscience}},
  volume={23},
  number={9},
  pages={3697--3714},
  year={2003},
  publisher={{Society for Neuroscience}}
}

@article{blais1997receptive,
  title={Receptive field formation in natural scene environments: comparison of single cell learning rules},
  author={Blais, Brian and Intrator, Nathan and Shouval, Harel and Cooper, Leon},
  journal={{Advances in Neural Information Processing Systems}},
  volume={10},
  year={1997}
}

\appendix

\section{Plasticity gradient with REINFORCE approximation}
Following the REINFORCE estimator~\citep{williams1992simple}, we approximate the gradient of the expected reward by
\begin{equation}\label{eq:app-1}
\nabla_\theta \langle R \rangle = \left\langle  (R - \bar{R}) \; \nabla_\theta \log \pi(\boldsymbol{\Delta}\mathbf{W} \mid \theta) \right\rangle.
\end{equation}
This results from applying the log-derivative trick on the expectation of Eq.~\ref{eq:app-1}
\begin{align*}
\nabla_\theta \langle R \rangle 
&=  \nabla_\theta \int \pi(\boldsymbol{\Delta}\mathbf{W} \mid \theta) \, R \, \de \boldsymbol{\Delta}\mathbf{W}  \\
&= \int \nabla_\theta \pi(\boldsymbol{\Delta}\mathbf{W} \mid \theta) \, R \, \de \boldsymbol{\Delta}\mathbf{W}  \quad \,\,\,\, \quad \quad \quad \, \quad \text{\color{cadetblue}{(Leibniz integral rule)}} \\
&= \int \pi(\boldsymbol{\Delta}\mathbf{W} \mid \theta) \, \frac{\nabla_\theta \pi(\boldsymbol{\Delta}\mathbf{W} \mid \theta)}{\pi(\boldsymbol{\Delta}\mathbf{W} \mid \theta)} \, R \, \de \boldsymbol{\Delta}\mathbf{W}  \\
&= \int \pi(\boldsymbol{\Delta}\mathbf{W} \mid \theta) \, \nabla_\theta \log \pi(\boldsymbol{\Delta}\mathbf{W} \mid \theta) \, R \, \de \boldsymbol{\Delta}\mathbf{W}  \quad \text{\color{cadetblue}{(log-derivative trick)}} \\
&= \left\langle  R \, \nabla_\theta \log \pi(\boldsymbol{\Delta}\mathbf{W} \mid \theta) \right\rangle_{ \pi} \\
&\approx \left\langle \underbracket[0.187ex]{(R - \bar{R})}_{\substack{\text{\color{carmine}{reward prediction}}\\[0.2ex]%
                \text{\color{carmine}{error}} }} \, \nabla_\theta \log \pi(\boldsymbol{\Delta}\mathbf{W} \mid \theta) \right\rangle .
\end{align*}
In the last expression we have introduced the baseline reward $\bar{R}$ as a control variate~\citep{lemieux2014control} commonly used for variance reduction of the expectation.
This heuristic uses the \textbf{reward prediction error} $\delta R = R - \bar{R}$ as a scaling factor for the direction of the update. This approximation assumes that $R$ is a smooth functional of $\mathbf{W}$ and that changes in $\theta$ affect $R$ primarily through their effect on the connectivity $\mathbf{W}$.

\section{Analysis of dynamics for each network}\label{apd:first}

\paragraph{Numerical solver.}
We located fixed points by solving for $\textbf{G}(\textbf{x})=\textbf{0}$ with
\[
\textbf{G}(\textbf{x}) = \textbf{x} - \mathbf{W} \phi(\textbf{x}) - \mathbf{W}_{\text{in}} \mathbf{u},
\]
using a damped Newton method. To avoid identifying duplicate fixed points, 
we initialized from $200$ random initial conditions and discarded solutions 
within a distance of  $\gamma \le 1e-5$ of one another. This resulted in a set of 
unique fixed points per input condition. 

\paragraph{Jacobian definition.}
We linearized the dynamics at the vicinity of each fixed point $\mathbf{x}^\ast$.  
The Jacobian is defined as
\begin{equation}
    \mathbf{J}(\textbf{x}^\ast) = -\mathbf{I}_N + \mathbf{W} \, \mathrm{diag}\!\left(1 - \phi^2(\textbf{x}^\ast)\right),
    \label{eq:Jct}
\end{equation}
and governs the local flow $\dot {\delta \textbf{x}} = \mathbf{J} \, \delta \textbf{x}$. 

\paragraph{Stability criteria.}
A fixed point is linearly stable if 
$\max \Re(\lambda(\mathbf{J})) < 0$. 

\paragraph{Eigenmode analysis.}
From the eigenvalues and eigenvectors of $J$, 
we extracted several diagnostics:
\begin{itemize}
    \item \textbf{Time constants and frequencies:} 
    Each mode with eigenvalue $\lambda = \beta + i\omega$ corresponds 
    to a decay time $-\tfrac{1}{\beta}$ (if $\beta < 0$) and an oscillation frequency $\omega / 2\pi$.
    \item \textbf{Non-normality:} Since the Jacobian eigenvectors are not orthogonal, small perturbations can have large transient effects even if all eigenvalue real parts are negative.
    Thus, we quantified the departure from normality with the \textbf{Henrici index}
    $\|\mathbf{J}\|^2_F - \sum_i |\lambda_i|^2$~\citep{murphy2009balanced,asllani2018structure}, which is indicative of transient amplification, where $\|\cdot\|_F$ stands for the Frobenius norm.
    \item \textbf{Transient gain:} We measured $\max_t \|e^{\mathbf{J} t}\|_2$ 
    on a grid of times $t$~\citep{christodoulou2022regimes, asllani2018topological}, capturing the degree of short-term amplification even when the dynamics are asymptotically stable.
    \item \textbf{Readout alignment:} 
    We computed the overlap of the corresponding readout vector $\mathbf{w}_{\text{out}}$ 
    with the eigenvectors of $\mathbf{J}$, to identify which eigenmodes directly influence network output.
    For each mode $i$, we characterise the alignment by
\begin{equation}
a_i = \bigl\|\langle \textbf{w}_{\text{out}}, \textbf{v}_i \rangle \bigr\|,
\end{equation}
where $\mathbf{v}_i$ are the normalized right eigenvectors of $J$. 
Large $a_i$ values indicate that the corresponding $i$-th mode contributes strongly to the output. 
In strongly non-normal systems we additionally verified projections using left--right biorthogonal eigenvectors.

    \item \textbf{Susceptibility:} We computed the linear response 
    $\mathbf{p} = (-\mathbf{J})^{-1} \mathbf{W}_{\text{in}}$,
    quantifying the steady-state sensitivity of each neuron to perturbations in the input.
\end{itemize}

\section{Tangent-propagation through learning}\label{apd:tangents}

\paragraph{Tangent-propagation through single trial time.}
To be able to compute the gradient of the weight updates with respect to the plasticity parameters, we propagate the gradients of the within-trial pre-activations $\mathbf{x}^t$, $\boldsymbol{\chi}_{k,\ell}^t \in \mathcal{R}^N$ (\textbf{state tangent}), the pre-activation trace $\bar{\mathbf{x}}^t$, $\boldsymbol{\psi}_{k,\ell}^t \in \mathcal{R}^N$ (\textbf{trace tangent}), and of the eligibility traces of each synaptic pair $ij$, $e^t_{ij}$, ${[\mathbf{z}^t_{k,\ell}]}_{ij}$ (\textbf{eligibility tangent}).
Thus we define the 
following within-trial tangents (sensitivities) with respect to the plasticity parameter $\theta_{k,\ell}$
\begin{align}\label{eq:single_sensitivities_1}
\boldsymbol{\chi}_{k,\ell}^t \dot{=} \frac{\partial \mathbf{x}^t}{\partial\theta_{k,\ell}},\qquad
\boldsymbol{\psi}_{k,\ell}^t \dot{=} \frac{\partial \mathbf{\bar{x}}^t}{\partial\theta_{k,\ell}},\qquad
\mathbf{z}_{k,\ell}^t \dot{=} \frac{\partial \mathbf{e}^t}{\partial\theta_{k,\ell}}.
\end{align}
We assume that the reward and baseline reward $R$ and $\bar{R}$ are not directly related to the plasticity parameters $\theta_{k,\ell}$, and thus we treat these variables and the reward prediction error as $\theta$-independent.

For convenience we define $\alpha=\mathrm{d}t/\tau$, and denote the derivative of the single neuron activation function with $\mathrm{d}\phi(x)=\phi^{\prime}(x)\,\mathrm{d}x$. The forward equations for these sensitivity parameters are 
\begin{align}
\boldsymbol{\chi}_{k,\ell}^{\,t+1} &= \boldsymbol{\chi}_{k,\ell}^{\,t} + \alpha\Big(-\boldsymbol{\chi}_{k,\ell}^{\,t} + \mathbf{W}^{(h)}\,\left(\text{diag}({\phi^{\prime}(\mathbf{x}^{t})})\cdot \boldsymbol{\chi}_{k,\ell}^{\,t}\right)+ \mathbf{U}_{k,\ell}^{(h)}\, \mathbf{r}^{t}\Big) \nonumber\\
\boldsymbol{\psi}_{k,\ell}^{\,t+1} &= \alpha_x\,\boldsymbol{\psi}_{k,\ell}^{\,t} + (1-\alpha_x)\,\boldsymbol{\chi}_{k,\ell}^{\,t+1},\nonumber\\[2pt]
\mathbf{z}_{k,\ell}^{\,t+1} &= \mathbf{z}_{k,\ell}^{\,t} + \mathrm{d}t
{ \;(\mathbf{\Delta x}^t)^\ell\otimes \left(\mathbf{r}^t\right)^k}\\
&\quad + \mathrm{d}t\!\sum_{\kappa,\lambda}\!\Big[
{\theta_{\kappa,\lambda}\,\lambda\,(\mathbf{\Delta x}^t)^{\lambda-1}(\boldsymbol{\psi}_{k,\ell}^{\,t}-\boldsymbol{\chi}_{k,\ell}^{\,t+1})\otimes \left(\mathbf{r}^{\,t}\right)^\kappa}
+ {\theta_{\kappa,\lambda}\,(\mathbf{\Delta x}^{t})^{\lambda}\otimes \kappa\,\left(\mathbf{r}^{t}\right)^{\kappa-1}\,(\text{diag}({\phi^{\prime}(\mathbf{x}^{t})})\cdot  \boldsymbol{\chi}_{k,\ell}^t)}
\Big], \nonumber
\end{align}
where $\text{diag}(\mathbf{y})$ denotes the matrix with $\mathbf{y}$ in the main diagonal, $\otimes$ denotes the outer product, while
\begin{equation}
\mathbf{U}_{k,\ell}^{(h)}\,\dot{=}\,\frac{\partial \mathbf{W}^{(h)}}{\partial\theta_{k,\ell}}
\end{equation}
stands for the inter-trial \textbf{weight matrix tangent}. The initial conditions for the three sensitivity parameters are zero at the beginning of each trial $\boldsymbol{\chi}_{k,\ell}^0 = \mathbf{0}$, $\boldsymbol{\psi}_{k,\ell}^0 = \mathbf{0}$, and $\mathbf{z}_{k,\ell}^0 = \mathbf{0} $.

At the end of each trial $h$ we have
\begin{equation}
\; \frac{\partial \boldsymbol{\mu}^{(h)}}{\partial \theta_{k,\ell}} = \eta\,\delta R^{(h)}\, \mathbf{z}_{k,\ell}^{T_h} \; .
\end{equation}

\paragraph{Propagating sensitivities through-learning (across trials).}
The derivative of the weights of trial $h+1$ w.r.t. $\theta_{k,\ell}$ accumulates the across trial sensitivities
\begin{equation}\label{eq:weight_sensitivities_1}
 \mathbf{U}_{k,\ell}^{(h+1)} \;=\; \mathbf{U}_{k,\ell}^{(h)} + \frac{\partial \boldsymbol{\mu}^{(h)}}{\partial \theta_{k,\ell}},\qquad \text{with   } \mathbf{U}_{k,\ell}^{(0)}=0.
\end{equation}
This sensitivity $\mathbf{U}_{k,\ell}^{(h)}$ couples back into the state tangent through the $\mathbf{U}_{k,\ell}^{(h)}\,\mathbf{r}^t$ term, which captures how $\theta_{k,\ell}$ affects later trials through the modified weights of earlier trials.

\subsection{Validation of weight updates gradients wrt plasticity parameters}
\begin{figure}[h!]
\includegraphics[width=\linewidth]{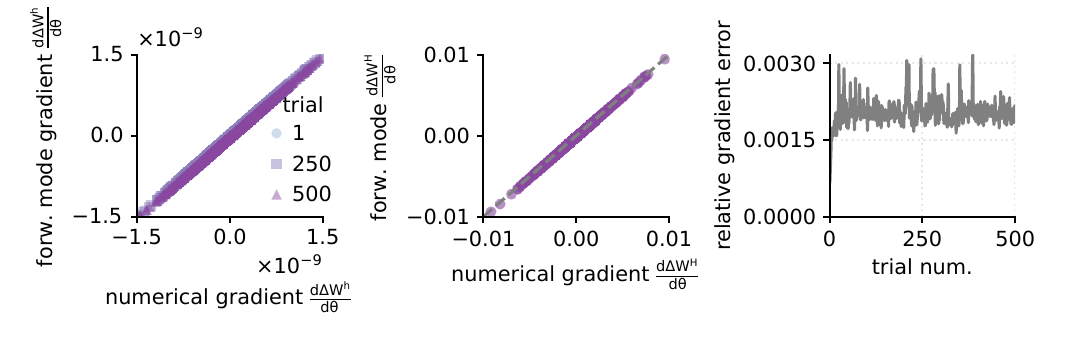}
\caption{\textbf{Validation of forward-mode gradient (FM) computation for the weight update wrt plasticity parameters $\theta$ against numerical gradients (FD).} \textbf{a.} Comparison of numerical gradient for per-trial weight update $\boldsymbol{\Delta} \mathbf{W}^{(h)}$ wrt plasticity parameter $\theta_{3,3}$ against the gradient obtained through forward mode differentiation for trials $1, 250, 500$ .  \textbf{b.} Comparison of cumulative gradient computed over $500$ trials for the weight update wrt plasticity parameters obtained numerically and through forward mode differentiation. The forward-mode differentiation provides an exact estimation of the plasticity update gradients.
\textbf{c.} Relative gradient error per trial computed as $\| {\frac{\de \boldsymbol{\Delta} \mathbf{W}}{\de \theta}}^{FM}  -{\frac{\de \boldsymbol{\Delta} \mathbf{W}}{\de \theta}}^{FD} \|  /  \| {\frac{\de \boldsymbol{\Delta} \mathbf{W}}{\de \theta}}^{FD} \|$. }
\label{app_fig:vali_eli}
\end{figure}

To validate the gradients wrt plasticity parameters obtained through forward mode differentiation, we compare both single-trial and multi-trial gradients (for $H=500$ trials) obtained with finite differences (FD) to those computed through forward mode differentiation (FM). To avoid observing discrepancies between the two versions of the gradients, we employ the same noise and environment inputs in all simulations. For this experiment we considered only $\theta_{3,3}=1$ nonzero, while all other entries of $\boldsymbol{\Theta}$ were zero. For the finite difference calculation, we run the training with $\theta + \epsilon$ and $\theta - \epsilon$ for $\epsilon=10^{-4}$ and approximate the gradient of the weight update with respect to the plasticity parameter as
\begin{equation}
\frac{\de \boldsymbol{\Delta} \mathbf{W}}{\de \theta_{k,\ell}}\approx\frac{\boldsymbol{\Delta} \mathbf{W}^+ \left(\theta_{k,\ell} + \varepsilon \right) - \boldsymbol{\Delta} \mathbf{W}^-\left(\theta_{k,\ell} - \varepsilon \right)}{2\varepsilon}.
\end{equation}
The resulting two versions of the gradients are in very close agreement throughout all $500$ trials (Fig.~\ref{app_fig:vali_eli}).

\end{document}